# MICRO-CRYSTAL GAAS ARRAY SUB-CELLS FOR SI TANDEM SOLAR CELLS

J.P. Connolly[1*], A. Nejim[3], A. Jaffré[1], J. Alvarez[1], J P. Kleider[1], D. Mencaraglia[1], Laurie Dentz[2], G. Hallais[2], F. Hamouda[2], L. Vincent[2], D. Bouchier[2], C. Renard[2]

[1]GeePs, Group of Electrical Engineering Paris, CNRS, CentraleSupelec, Université Paris-Saclay, Sorbonne Université, 3&11 rue Joliot-Curie, Plateau de Moulon, 91192 Gif-sur-Yvette CEDEX, France

[2]C2N, Centre de Nanosciences et de Nanotechnologies, CNRS, Université Paris-Saclay, 10 Bd Thomas Gobert, 91120, Palaiseau, France

[3]SILVACO Technology Centre, Compass Point, St. Ives, Cambridgeshire PE27 5JL, UK

ABSTRACT: This work reports optical and electronic numerical modelling of a novel emerging structure which is the GaAs nanocrystal on Si tandem solar cell by epitaxial lateral overgrowth, a technique which allows defect free material growth. The techniqueconsists of creating nucleation sites in a silicon surface SiO$_2$ layer and initiating growth of nanoscalescale seeds, whereby strain energy remains below the Matthews-Blakeslee strain relaxation limit. This leads to Al$_x$GaAs growth in micro-crystals without generation of material defects. The focus of this presentation is optical and electrical modelling of nanocrystals for applications in the very active field of silicon based multijunction solar cells, and design of a Al$_x$GaAs/Si two terminal tandem, for compositions ranging from x=0 to x=30% in absorber layers. We present a model of the complete structure in two dimensions, consisting of a Al$_x$GaAs high bandgap subcell connected with a tunnel junction to the low bandgap Si junction. The elaboration of models is described, with an emphasis on the Al$_x$GaAs crystal featuring a non-planar pn-junction, and a focus on the optical properties of this lattice of micrometric AlGaAs crystals and in particular their light trapping properties from the resulting surface texture. The question of Al$_x$GaAs surface coverage is addressed, given that neighbouring Al$_x$GaAs crystals have different crystal orientations on a (111) Si surface, such that any coalescence of neighbour Al$_x$GaAs crystals leads to crippling defects at their interface. The result is that some high energy incident light above the Al$_x$GaAs bandgap is nevertheless transmitted directly to the Si cell, such that the resulting photogenerated carriers thermalise to the Silicon bandgap, and result in a loss of efficiency. The interface between Al$_x$GaAs and Si subcells is addressed, with an emphasis on current transport efficiency through the nanoseeds and tunnelling currents through appropriately designed SiO$_2$ buffer layers. This work therefore presents a theoretical framework for evaluating the potential of Al$_x$GaAs nanocrystal growth on Si for light trapping, for GaAs silicon two terminal tandem cell performance including tunnel junctions, and provides models and design rules for efficient Al$_x$GaAs microcrystal arrays as high bandgap subcells for tandem solar cells on silicon.

Keywords: III-V, silicon, texturing, epitaxial lateral overgrowth.
Corresponding author : james.connolly@centralesupelec.fr

1 Introduction

The multijunction solar cell is the most successful high efficiency concept. This has led to much work on on this topic based around silicon, for economic and industrial reasons [1].

As part of these efforts, there has been significant effort in integration of III-V semiconductors on Si substrates for photovoltaics and more broadly in opto-electronics. Avenues followed involve a range of techniques [2] which generally need to manage defect densities to lead to usable devices. One technique suggested some decades ago [3] which avoids these issues is the epitaxial lateral overgrowth (ELO) method which presents significant practical growth and fabrication challenges in obtaining high quality III-V films.

We explore the application of these micro-crystal arrays as the high band-gap subcell of III-V / Si two terminal series connected tandem solar cells. We evaluate strategies for optimum tandem designs in light of limiting efficiencies.

We focus on two topics of interest, which are the increased light-matter interaction resulting from light scattering by the micro-crystal array, and the monolithic integration and efficient current transport in the resulting III-V / Si multilayer photovoltaic device. The result is an innovative design which benefits from built-in texturing and light trapping, while developing a promising technological route for III-V integration on Si for wider applications.

2 Experimental context

This paper builds on the recent work [3] which has demonstrated technologically attractive and high quality arrays of single crystal GaAs micro-crystal arrays on Si by epitaxial lateral overgrowth (ELO). As illustrated in figure 1, ELO growth starts with a Si surface on which a 2nm surface SiO$_2$ layer is fabricated. Nucleation sites of diameter ≈50nm are etched in this oxide. Growth is initiated by epitaxial methods in these nucleation sites.

This technique has allowed the growth of defect free Al$_x$GaAs crystals on Si despite the mismatched Al$_x$GaAs/Si atomic lattices because the small contact area and 3D growth mode ensures strain energy never exceeds the Matthews-Blakeslee limit [4], and that no strain relaxation and formation of lattice defects occurs. Current growth methods yield two crystal types. The first is flat rectangular crystals with height to width ratios of 1/4, and of dimensions from 1μm to 2μm wide shown in figure 2(a). The second consists of the same rectangular base with a hexagonal "cap" with facets fixed at 30° from the horizontal, as shown in figure 2(b), which we call "textured".

The materials available are Al$_x$GaAs with x=0-40% for absorbers and higher x=75 for thin layers including windows or the front emitter.

We note that Al$_x$GaAs is a complex material. The Al content leads to instabilities in particular with respect to reaction with oxygen such that $Al_xGaAs$ layers need to be isolated from the atmosphere. This isolation is provided here by contact transparent conducting oxide layers, and anti-reflection coatings.

We note also intrinsic materials problems, in particular the well known DX centre [5] which is associated with Al compositions greater than ≈22% and high doping, which is nevertherless not a critical issue in the Al fractions available, and given the low doping we use in the absorber layer as detailed below.

Finally, GaAs at the lower x=0 composition has long been a nearly ideal solar cell material achieving to date AM1.5G efficiencies of 29% [6], close to the radiative limit of 31% [7].

This concludes the definition of available materials, for which we next present simulations evaluating potential device performance.

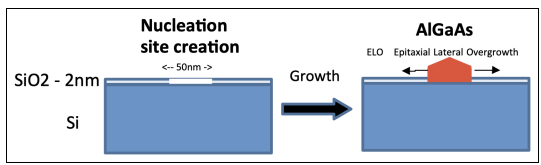

**Figure 1** : the epitaxial lateral overgrowth technique, whereby AlGaAs nucleations sites are created by opening holes of some tens of nanometres in a two nanometre thick surface oxide.

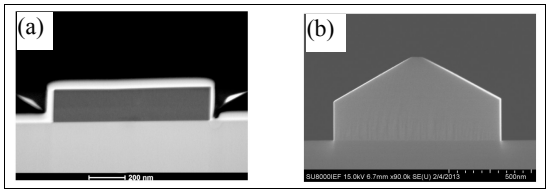

**Figure 2** examples of 1μm wide (a) flat and (b) faceted ELO crystals

3 Theoretical limits and device design

It is worth reminding the well-known radiative efficiency limits with familiar efficiency – bandgap contour plots [1] which give an appreciation of potential efficiencies of proposed designs. The fundamental efficiency limit for a silicon based tandem is 41.9% under standard test conditions (STC), for of 1.74eV (top) and and 1.12 (bottom Si).

3.1 Optimum efficiency higher gap top cell

The optimum Si-based tandem upper gap of 1.74eV corresponds to $Al_{25}Ga_{75}As$. This is a material with sufficienty materials properties for solar cells, as we have seen just above. The material is just above the range where materials issues start to become significant at x≈22% but for highly doped materials. Since we are considering low-doped materials for the absorber region with a large depletion region with field driven transport, this composition is well within the tolerances for efficient photogenerated carrier diffusion and collection.

We can in addition retain higher composition $Al_xGaAs$ for the frontmost emitter layer. This is because the role of this layer is to set up the *pn* junction and internal field. For this role, poor minority carrier transport is not an obstacle and indirect absorption is an advantage since light mainly absorbed in the more efficient absorber region.

We will therefore in further sections evaluate the potential performance of $Al_{25}GaAs$ microcrystals.

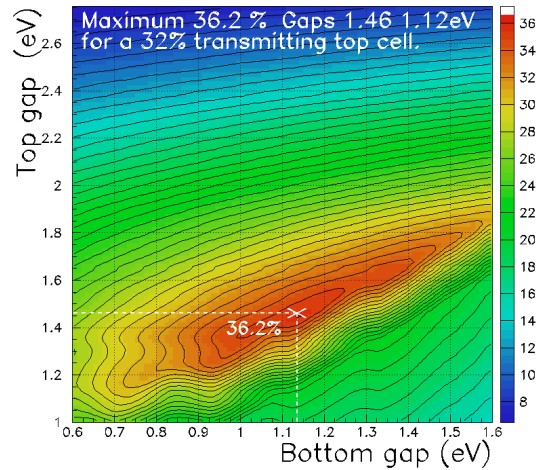

**Figure 3** Radiative efficiency limit (STC) of a tandem cell where the top cell is optically thinned across the wavelength range such as to achieve an ideal tandem cell which features Si as the bottom cell at 1.12eV. The efficiency maximum is found a touch above GaAs for $Al_2Ga_{98}As$. This can be simplified by substituting GaAs which barely changes the achievable efficiency which for GaAs is 36%.

3.1 Optimum transport lower gap top cell

While the gap of GaAs is too low for a current-matched tandem cell on Si, with the GaAs cell over-producing, one can optically thin the top cell to achieve current matching.

This is non-ideal and reduces potential efficiency because photons with energies above the GaAs gap are absorbed in the Si. The resulting minority carriers are generated with higher energies than necessary, and lose this energy in thermalising to the Si band-edge. This may however be advantageous for a tandem if the thermalisation loss is small enough to allow high efficiency, and if tandem design is otherwise facilitated, as it is here as we shall see.

The potential efficiency can be evaluated by evaluating limiting radiative efficiency of an non-opaque or optically transmitting top cell which does not absorb all the light above its bandgap.

We tune the transmission of such a top cell such that the optimum bottom bandgap corresponds to Si. This procedure yields an optimum for a non-opaque top cell which transmits 32% of light above its bandgap. This has an achievable efficiency of 36.2% for al $Al_2Ga_{98}As$ cell on Si, which is essentially indistinguishable from a thinned GaAs/Si tandem.

We conclude with an important point which is that this "optical thinning" can be achieved by physically thinning the top cell material, or by depositing the top cell with partial coverage, leaving sections of the bottom Si cell directly exposed to the incident spectrum.

This solution is ideally suited to the ELO Al$_x$GaAs cell growth, since growth proceeds by nucleation of independent crystals. Complete coalescence is in fact a challenge, and while it may be in principle achieved, independent crystals are technologically far easier to fabricate.

For completeness, we mention here work towards complete coverage in similar work by Strömberg *et al.* [8]. This investigates ELO fabrication of GaAsP on Si.

## 4 Simulations

The simulations first aim to analyse the fabricated devices described in section 2 in order to establish the potential performance of available materials. We start by briefly summarising the modelling strategy, before presenting the properties of the structure assumed in the modelling, before presenting results of electrical and optical modelling.

### 3.1 Model specification

The simulation in this work is carried out with Silvaco finite element numerical software in the Victory suite, in particular the process, meshing, and device simulation modules which we will not describe and instead refer the reader to manuals and publications which provide these details [9].

The complexity of the multilayer textured structure consisting of two different cells and a range of materials, and the dimensions of the smallest elements in principle require a precise finite difference time domain model (FDTD). This is however computationally expensive. Furthermore the optically relevant layers range from 1μm to hundreds of microns, and the refractive index contrast with the thin Al$_x$GaAs emitter layer at the front remains small. For these reasons, these studies rely on ray tracing which is sufficient for preliminary investigations, subject to FDTD studies in future work if necessary.

|  | Dimension (μm) | Material | doping |
|---|---|---|---|
| Transparent conducting oxide | 0.1 | ITO | - |
| **Top cell** | | | |
| Crystal facet angle | 30° and 0° | - | - |
| Emitter | 0.1 | Al$_x$GaAs, x=0.4-0.7 | p 1E18 |
| Base | 1 - 2 | Al$_x$GaAs, x=0.3-0.4 | n 1E15 |
| **Interface oxide** | | | |
| Oxide | 2E-3 | SiO$_2$ | - |
| Seed | 0.1 | Al$_x$GaAs, x=0.3-0.4 | n 1E15 |
| **Silicon bottom cell** | | | |
| Tunnel 1 | 0.1 | Si | n 1E19 |
| Tunnel 2 / base | 0.1 | Si | p 1E19 |
| Wafer | 2 - 250 | Si | n 1E15 |
| Emitter/back contact | 0.1 | Si | n 1E20 |

Table 1 model device specification range of materials parameters consistent with current fabrication techniques.

### 3.2 Device specification

Since we are interested in evaluating the current materials, modelled devices are not optimised from the perspective of layer dimensions, doping, and light interaction (anti-reflection (AR) coatings) in order to match the current experimental achievements.

The modelled device specifications are detailed in table 1. They include both flat and textured geometries with the doping, dimensions indicated, the nucleation site. The chosen compositional parameters are an Al$_{40}$Ga$_{60}$As emitter and Al$_{30}$Ga$_{70}$As absorber.

Also indicated are preliminary definitions of the tunnel layers and the silicon bottom cell which are however not simulated in this work, again in line with prioritising the experimental status.

Figure 4 shows succinctly the modelled structures, flat and textured, which are analogies to the experimental structures in figure 2. The model structures differ in the presence of a contact transparent conducting oxide and a front surface emitter layer which are in development experimentally.

Figure 5 completes the numerical description of the full cell ranging from the nanometre scale nucleation site to the hundred micron scale silicon substrate. This study concentrates on the top cell and does not consider the tunnel layers and Si performance which are to be implemented experimentally once top cell devices are operational.

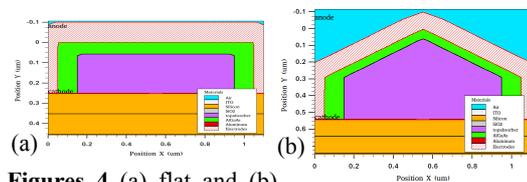

**Figures 4** (a) flat and (b) textured GaAs crystals as defined by the process model for the lower 1μm dimension and inter-crystal separatin of 100nm. Also show are the tunnel junction layers for completeness which are not implemented in this study.

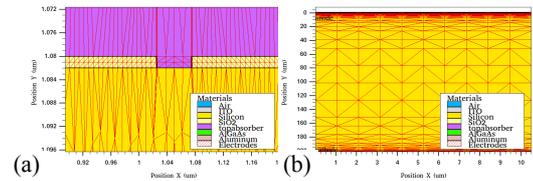

**Figure 5** Schematics of the device and numerical Delaunay mesh over the range of scales from (a) the nanometre scale nucleation site to (b) the full device scale which is dominated by the Si substrate at the hundred micron scale.

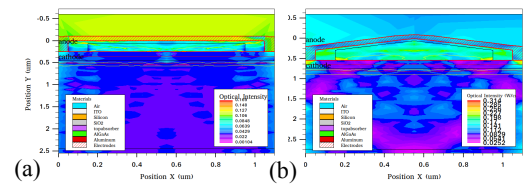

**Figure 6** Light intensity map in cross section of (a) a flat 1μm crystal and (b) the textured analogue

### 3.3 Light management

Figure 6(a) shows light intensity maps in cross section of the flat device model. We note some light diffration at the edges due to the crystal separation. The centre shows an minor artefact corresponding to the nucleation site which however should not be considered reliable given the ray tracing model applied, and the size of this feature well below the incident photon wavelengths. Figure 6(b) shows the analogue for the textured sample.

We note first the difference in light intensity above the cell, with the flat structure reflecting significantly more light than the textured. This confirms the role of texturing as improving light interaction by reducing surface reflectivity, a well known phenomenon used in textured Si cells as pioneered by Green [10].

Quantification and optimisation of this reduction of reflectivity is at this stage preliminary since we are not considering AR coats at this stage as mentioned previously.

The second effect demonstrated at this state is light refraction within the cell by the surface texture provided by the micro-crystal array. This is evident in the light refraction visible in firure 6b compared to 6a. In the flat case, features in light distribution are visible laterally but as mentioned just above but are due to the presence of spacers.

In the textured case we see enhanced light intensity below the microcrystal which leads to enhanced photogeneration at shorter depths in the Si cell.

This reproduces the same light refraction and light trapping phenomena which are now standard in high efficiency textured Si solar cells[10].

We conclude that the $Al_xGaAs$ ELO crystal array provides the same light management design features which allow a thinning of Si substrates and a higher efficiency due to shorted diffusion scales for carrier collection in the Si. They also provide similar advantages in the $Al_xGaAs$ higher gap subcell.

Both these phenomena will be quantified in future work, including optical coupling between the two. On that point, we note that $Al_xGaAs$ and Si have very similar real refractice indices in the wavelength region below the $Al_xGaAs$ absorber bandgap range [11].

| Design | $J_{SC}$ ($mA/cm^2$) | $V_{OC}$ ($V$) | FF (%) | Efficency (%) |
|---|---|---|---|---|
| Flat | 61.9 | 1.38 | 88 | 7.58 |
| Textured | 91.5 | 1.39 | 90 | 11.5 |

**Table 2** Simulated device performances for material geometries fabricated to date which lacking optimisation feature low efficiencies, but with textured devices close to twice the flat device performance.

3.4 Device performance

We first present in table 2 the performance of current materials with, as noted above, an $Al_{40}Ga_{60}As$ emitter and $Al_{30}Ga_{70}As$ absorber. Lacking optimisation of structure and materials, these are projected to achieve low efficiencies as might be expected.

More important is to note the significant efficiency enhancement in the textured device which at 11.5% is close to twice the efficiency of the flat at 7.6%. This is due for a number of contributing factors.

The difference in absorbing volume is the first. The thin cell, 1μm across and only 0.25μm high, is too thin to absorb the incident spectrum efficiently. In addition, surface reflection is particularly high in the absence of an AR coat. Finally, carrier diffusion lengths are maximised by the 1/4 hight to width geometry and the imposition of current transport through the centred nucleation site.

The textured device, in contrast, benefits first from a greater thickness from the hexagonal cap which raises the maximum thickness by about a third, and increases absorption. This is amplified by the surface refraction by the facets which adds light trapping in the top cell, increasing the absorption of the greater absorbing thickness compared to the flat case. Finally, of course, the surface reflection is reduced by the texturing as we have seen earlier in the light management analysis.

We do not include here the quantum efficiencies and light current characteristics for succinctness but note they are shown in the EUPVSEC2025 presentation available on the conference website.

We conclude by evaluating routes to improvement for the top cell. The first is the modification of the emitter by increasing the Al composition to $Al_{70}Ga_{30}As$, close to the maximum, and reducing the absorber Al composition zero, that is, GaAs.

We also eliminate the spacer which is currently not feasible but which is under investigation by other workers in similar studies [7].

The third and final modification is to increase crystal width to the current achieved of 3μm. This gives a base height of 0.75μm and a cap height of approximately 1.3μm, as shown in figure which is sufficient for complete light absorption above its bandgap of 1.42eV. if slightly above the largest size achieved to date which is 2μm.

The quantum efficiency and power/voltage curve for this device are shown in figures 8a and 8b. This device achieved efficiency f 22%, for a JSC of 216mA/cm2, VOC of 1.15V, and fill factor of 89%.

Extrapolating this to our target 32% transmitting top cell yields a final top cell efficiency in this simple first case of 15.0%.

The achievable tandem efficiency remains low with a Si bottom cell efficiency of just 9.2% based on an optically coupled model for which we show the EQE only for brevity in figure 9. This uses a 200μm Si cell optically coupled to a GaAs surface microcrystal array with 32% transmission. The combined tandem efficiency is of 24.2%, which remains low since as we have already mentioned the intention of this presentation is a preliminary evaluation of potential performance rather than a full optimisation which will follow.

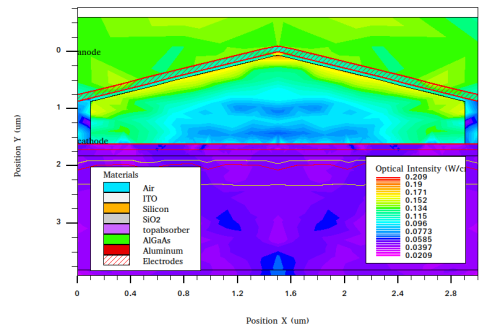

**Figure 7** Higher efficiency design consisting of GaAs cell with no spacer, and 3μm width leading to a crystal base height 0.75μm and cap height 1.55μm sufficient for total absorption in the GaAs.

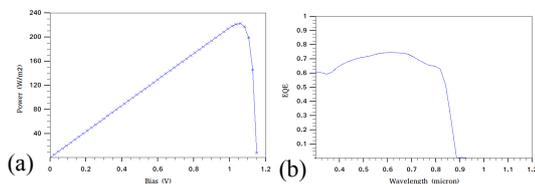

**Figure 8** power-voltage (a) and quantum efficiency (b)

of idealised device with no spacer and a GaAs top composition

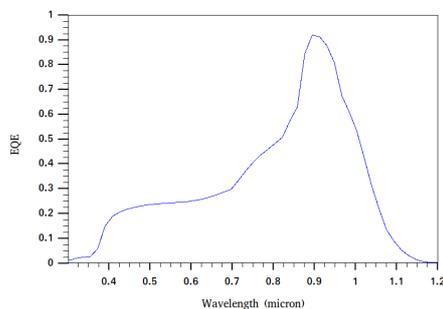

**Figure 9** Quantum efficiency of a 200μm think Si bottom cell under 32% optically transmitting GaAs microcrystal array for current matching.

5 Conclusions

We have presented modelling of textured ELO AlGaAs crystals on Si. We find first that these structures allow light scattering of the same type as used in high efficiency standard silicon solar cells. An array of AlGaAs microcrystals is therefore very well suited to provide an effective texturing on a tandem III-V/Si solar cell, with the Si cell consisting of a thinner Si substrate as in high effiicency Si solar cells.

We have evaluated the potential efficiency of the current state of the art ELO crystals. Efficiencies remain low because of a lack of optimisation of optical and structural properties.

We have shown that based on current materials a 32% transmitting top GaAs cell with an efficiency of 15% is achievable, consistent with current matching to a lower gap Si cell.

This is projected to reach efficiencies in tandem structures which reach 24%, which while not impressive for a complex tandem device is nevertheless consistent current materials status and where identified routes to significantly higher performance is clear, and consists of standard solar cell optimisation.

We also note that a higher bandgap design has been identified as an $Al_{20}Ga_{80}As$ providing current matching. While this is currently not achievable since it requires complete coverage, work in the field towards complete top cell coverage makes this an even higher efficiency design to consider in future.

The next steps will be to simulate the top and bottom cell structures and combined tandem performance with experimental developments in the direction of top cell, bottom cell, and complete tandem devices, and thereby evaluate the light management and tandem potential of ELO AlGaAs for photovoltaic and broader applications.

Acknowledgments

The authors acknowledge the support of the French National Research Agency (ANR) financing the project HELLO-PV (ANR22-CE050-0011) which has made this work possible.